\pdfoutput=1
\documentclass[epsf,aps,preprint,showpacs,preprintnumbers,amsmath,amssymb,footinbib]{revtex4}
\usepackage{bm}
\usepackage[pdftex]{graphicx}
\usepackage[colorlinks=true, pdfstartview=FitV, linkcolor=red, citecolor=blue, urlcolor=blue]{hyperref}
\newcommand{\beq}{\begin{equation}}
\newcommand{\eeq}{\end{equation}}

\def\renlam{\Lambda_{\rm ren}}
\def\wtepert{\widetilde{E}_0^{\rm pert}}
\def\epert{E_0^{\rm pert}}
\def\enon{E_0^{\rm non}}
\def\mdebye{m_{\rm Debye}}

\def\U1{\text{U}(1)}

\def\renzR{{\cal Z}_{\cal R}}

\def\epjc#1#2#3{Eur.\ Phys.\ Jour.\ C{\bf #1}, #2 (#3)}
\def\epl#1#2#3{Eur.\ Phys.\ Lett. {\bf #1}, #2 (#3)}
\def\ibid#1#2#3{{\it ibid.} {\bf #1}, #2 (#3)}
\def\ijm#1#2#3{Intl. Jour. Mod. Phys. {\bf #1}, #2 (#3)}
\def\jhep#1#2#3{Jour. High Energy Phys. {\bf #1}, #2 (#3)}

\def\npb#1#2#3{Nucl. Phys. B {\bf #1}, #2 (#3)}

\def\plb#1#2#3{Phys. Lett. B {\bf #1}, #2 (#3)}

\def\prd#1#2#3{Phys. Rev. D {\bf #1}, #2 (#3)}
\def\prl#1#2#3{Phys. Rev. Lett. {\bf #1}, #2 (#3)}
\def\phr#1#2#3{Phys. Rep. {\bf #1}, #2 (#3)}

\begin{document}
\preprint{KUNS-2220}
\title{Zero Point Energy of Renormalized Wilson Loops}
\author{Yoshimasa Hidaka$^{a}$
and Robert D. Pisarski$^{b}$}
\affiliation{
$^a$Department of Physics, Kyoto University, Sakyo-ku, Kyoto 606-8502, Japan\\
$^b$Department of Physics, Brookhaven National Laboratory, Upton, NY 11973, USA\\
}
\begin{abstract}
The quark antiquark potential,
and its associated zero point energy, can be extracted from lattice
measurements of the Wilson loop.
We discuss a unique prescription to renormalize
the Wilson loop, for which the 
perturbative contribution to the zero point
energy vanishes identically.  A zero point energy can arise
nonperturbatively, which we illustrate by considering effective
string models.  The nonperturbative contribution to the
zero point energy vanishes in the Nambu
model, but is nonzero when terms for
extrinsic curvature are included.  At one loop order, the nonperturbative 
contribution to the zero point energy is
negative, regardless of the sign of the extrinsic curvature term. 
\end{abstract}
\date{\today}
\maketitle

\section{Introduction}
\label{sec_intro}

The Wilson loop has a privileged status in a gauge theory.
Although a nonlocal and composite operator, 
with dimensional regularization
smooth loops in $3+1$ dimensions are rendered finite by
the usual renormalization constants for the gluon \cite{renloop_pert}.  

Lattice regularization, however, introduces
an additional divergence. Considering
the Wilson loop as the propagator for an infinitely heavy test quark,
the new divergence, $E_0$, represents
mass renormalization for the test quark.  
$E_0$ has dimensions of mass, so if ``$a$'' 
is the lattice spacing, $E_0 \sim 1/a$, and multiplies the length of the loop.
Extracting the quark
antiquark potential from the Wilson loop in the usual manner,
$E_0$ corresponds to the zero point energy of the potential at asymptotically
large distances. Although the
(bare) quark antiquark potential has been measured with precision
\cite{lu_term,lattice_potential,lu_wz,teper}, $E_0$ 
is usually ignored. 

It is thus of interest to know how to renormalize $E_0$ 
on the lattice.  This is especially important for thermal Wilson
loops, where in a pure gauge theory the Polyakov loop
is the order parameter for deconfinement.
While the bare Polyakov loop vanishes in the continuum limit,
the renormalized loop does not
\cite{ren_lat1,ren_lat2,ren_lat3,ren_lat4,ghk,gava_jengo,heller,lattice_effective,loop_phen}.
However, as $E_0$ varies, the
the renormalized Polyakov loop $\ell$ changes as
$\ell \rightarrow {\rm e}^{-E_0/T} \ell$;
see, {\it e.g.}, appendix C of ref. \cite{ghk}.
This nonperturbative ambiguity cannot be fixed by appealing
to the perturbative regime at high temperature.  

In the first section of the paper, we follow previous analysis
\cite{ren_lat1,ren_lat2,ren_lat4,ghk} and suggest that there is a natural
way to renormalize the Wilson loop such that 
the perturbative contribution to the zero point
energy vanishes identically, $\epert = 0$.  

This does not exclude nonperturbative contributions to the zero point
energy.  We illustrate this by considering $\enon$ in one possible
model of confinement, that of string models.
If $\sigma$ is the string tension, then just on dimensional grounds,
$\enon \sim \sqrt{\sigma}$ is possible.
Even so, in the Nambu model \cite{alvarez,arvis}, or variants thereof
\cite{lu_wz,pol_strom,drummond}, $\enon = 0$.  

A nonzero value of $\enon$ is generated only by 
models with massive
modes on the world sheet.  This arises by adding
terms for the extrinsic curvature of the world
sheet.  A simple computation shows that at one loop order,
$\enon$ is nonzero and negative, whether
the coupling for the extrinsic curvature term is positive
\cite{rigid1,rigid2,rigid3,rigid4}
or negative \cite{negative,conf1,conf2,conf_temp}.  

On the lattice, numerical simulations find that
the quark antiquark potential appears to agree well with the simplest
Nambu model down to rather short distances \cite{lattice_potential}.
This suggests that on the world sheet, any massive modes are heavy.
In this case, the
effective string theory is strongly coupled, so that the results of
a one loop computation are only suggestive.

\section{Renormalized zero point energy}
\label{ren_wilson}

Consider a rectangular Wilson loop of length 
$t_\text{tot}$ and width $R$.  When
$t_\text{tot} \gg R$, 
the vacuum expectation value of the bare Wilson loop can be used
to define the quark antiquark potential, 
\begin{equation}
\langle {\cal W} \rangle =
\left\langle 
\exp\left( i g \oint_{\cal C} A_\mu 
\; d x^\mu  \right) 
\right\rangle 
= \exp\left( - V(R)\; t_\text{tot} \right) \; .
\label{wilson_loop}
\end{equation}
At large distances, 
\begin{equation}
V(R \rightarrow \infty) 
\sim \sigma \, R + E_0 - \frac{\alpha}{R} + \ldots
\; \label{zero_point_energy}
\end{equation}
Here $\sigma$ is the string tension, $E_0$ is the zero point energy,
and ``$\alpha$'' is a constant.  

Assume that the theory has no dynamical quarks and confines, with the
loop in the fundamental representation.  Then
$\sigma \neq 0$, and $\alpha$ 
is universal, $= \pi/12$ in four spacetime dimensions \cite{lu_term}.
In quantum mechanics, the value of the zero point energy
simply produces 
a phase which multiplies the wave function, and is of no physical
consequence.

The potential above is a bare quantity, but 
only the zero point energy is ultraviolet divergent.
The Wilson loop is related to the propagator for a massive test quark,
with the zero point energy the additive shift in the mass.
As the mass of the test field goes to infinity,
the integral for this mass divergence is over three, 
instead of four, dimensions:
\begin{equation}
\epert \sim  - \;
\; g^2 \int^{1/a} \frac{d^3k}{(2\pi)^{3}} \; \frac{1}{k^2} 
+ \ldots \; 
= - \;  \; c_1 \; g^2 
\left( 1 + c_2 g^2 + \ldots \right) \; \frac{1}{a} \; .
\label{eq1}
\end{equation}
The constants $c_1$, $c_2$, 
{\it etc.} depend upon the representation of the test particle,
the details of lattice discretization, {\it etc.}

The possibility of a zero point energy was recognized when
the renormalization of the Wilson loop was first considered
\cite{renloop_pert}.  In perturbative computations it is of little
consequence, because then dimensional regularization is natural,
and the integral in 
Eq.~(\ref{eq1}) automatically
vanishes, as a purely powerlike divergence in an odd number
of dimensions.  This remains true with 
other gauge invariant regulators, such as higher derivatives or
Pauli-Villars, since they also eliminate such power law divergences.

To define a renormalized loop on the lattice, we introduce the
renormalization constant $\renzR$, which is a function of the
(bare) coupling, $g^2$:
\begin{equation}
{\cal W}^{\rm bare}_{\cal R}  =
\renzR(g^2)^{L/a} \;
\; {\cal W}^{\rm ren}_{\cal R} \; .
\label{appendix3}
\end{equation}
With $L$ is the length of the loop, $L/a$ is the number of links
for the path.  

One way of extracting the renormalization constant $\renzR(g^2)$
is to compute for
different values of the lattice spacing, $a$,
holding the physical length, $L$, fixed \cite{ren_lat2,ghk}.  
This can be done directly from the numerical simulations, without
using the perturbative expansion of Eq.~(\ref{eq1}).

Another way of computing $\renzR(g^2)$ is to compare the
potential at short distances, $V(R)$ as $R \rightarrow 0$, to the result
in perturbation theory \cite{ren_lat1,ren_lat4,ghk}.  
In a pure $SU(3)$ gauge theory, the two methods agree with one another
to the numerical accuracy tested \cite{ghk}.  

Under this renormalization, it is possible to redefine
\begin{equation}
\renzR(g^2)
\rightarrow \renzR(g^2)^{L/a} \;
{\rm e}^{- \wtepert L}/{\cal Z}_0\; ,
\label{change1}
\end{equation}
so that the renormalized Wilson loop becomes
\begin{equation}
{\cal W}^{\rm ren}_{\cal R} \rightarrow
\; {\rm e}^{+ \wtepert L} \; {\cal Z}_0 \; 
{\cal W}^{\rm ren}_{\cal R} \; .
\label{change2}
\end{equation}

In these expressions 
${\cal Z}_0$ is a pure number, which
just shifts the overall normalization of the loop.
We can always choose this normalization by considering very small
loops; for Polyakov loops, this corresponds to 
the limit of very high temperature.  For small loops, perturbative
corrections are computable as a power series in
$\sim g^2(L) \sim 1/\log(L)$.  We can thus eliminate
${\cal Z}_0$ by requiring that the small loops, suitably
normalized, approach unity as $L \rightarrow 0$.

Less trivial is the change due to $\wtepert$, which shifts the renormalized
loop by $\exp(\wtepert L)$.  This cannot be eliminated considering
small loops, $L \rightarrow 0$, as $\wtepert$ represents a 
correction in a power of $L$, which in an asymptotically free theory
is nonperturbative.

We suggest that under perturbative renormalization, the only consistent
choice is to take $\wtepert = 0$.  We can 
renormalize the loop at zero temperature, where gluons are massless.
For massless fields, the linearly divergent integral is uniformly 
proportional to the ultraviolet cutoff, which is $\sim 1/a$.  The basic
point is that there are no finite terms $\sim a^0$.

This depends crucially upon the fact that at zero temperature, gluons
are massless order by order in perturbation theory.  
Consider how the integral in Eq.~(\ref{eq1}) changes if the
gluons did have a mass, $m$:
\begin{equation}
\epert(m)
\sim - \; g^2 \; \int^{1/a} \frac{d^3k}{(2\pi)^{3}} \;
 \frac{1}{k^2 + m^2} 
= - \; g^2 \left( c_1 \; \frac{1}{a} - \;
e_2 \; m + e_3 \; m^2 \; a + \ldots \right) \; ,
\label{eq2}
\end{equation}
for some constants $e_2$, $e_3$, {\it etc.}  The ultraviolet divergent
term, $\sim 1/a$, is independent of the mass $m$, and so the coefficient
$c_1$ is the same as in Eq.~(\ref{eq1}).  When $m\neq 0$, though,
there is a term $\sim e_2 m$ which contributes a finite amount to
the zero point energy.  There are also terms at higher order which
vanish as $a \rightarrow 0$, $\sim e_3 m^2 a$, {\it etc.}

Our point is simply that when $m=0$, then
$e_2 = 0$, and there is no contribution
to the zero point energy.  This can be made more general.
Consider first a non-Abelian gauge theory without dynamical quarks,
and let the renormalization mass scale of the theory be $\renlam$.
To be invariant under the renormalization group,
$\wtepert$ can only depend upon $\renlam$ as
$\wtepert \sim \renlam
\exp(-\int dg/\beta(g^2))$, where $\beta(g^2)$ is the $\beta$-function
for the theory.  Doing so, however, means that 
$\wtepert$ is a dynamically generated
mass scale.  Such a nonperturbative mass scale is inconsistent
with perturbative renormalization.

Strictly speaking, this argument fails in the presence
of dynamical quarks, where one could have 
$\wtepert \sim m_{\rm quark}$, with $m_{\rm quark}$ a current
quark mass.  We suggest that $\wtepert$ vanishes even in the presence
of dynamical quarks.  
However quarks modify the
gluon propagator, the perturbative contribution to
the zero point energy arises from
a linearly divergent integral over a massless field, as it does in the
pure glue theory.

The bare zero point energy,
Eq.~(\ref{zero_point_energy}), is then a sum of two contributions,
\begin{equation}
E_0 = \epert + \enon \; .
\label{zero_sum}
\end{equation}
The perturbative contribution is ultraviolet divergent, with
$a \epert \neq 0$ as $a \rightarrow 0$.  This does not exclude
nonperturbative contributions, $\enon$, for which $a \enon \rightarrow 0$
as $a \rightarrow 0$.  We compute $\enon$ in string models in the 
next section.  Our point is simply that $\epert$ can be
uniquely determined: there is no freedom to vary $\epert$ by a finite
amount by introducing $\wtepert$.

In particular, this implies that it is not possible to choose a $\wtepert$
so that the renormalized potential, $V(R)$ vanishes at a given distance
\cite{ren_lat3}.  Of course the renormalized potential will vanish at
some distance $R_0$, but this cannot be fixed {\it a priori}.  Instead,
if follows from the renormalized quark antiquark potential.

There are several notable examples 
when the gluon develops a ``mass'', and a finite zero point energy arises
perturbatively.   In all of these
cases, however, renormalization proceeds as for $m=0$; there is
no ambiguity in how to compute such terms.

The first is at nonzero temperature, where a Debye mass
$\mdebye \sim g\, T$ arises for $A_0$.  Comparing to 
Eq.~(\ref{eq2}), this generates a zero point energy
$\sim g \, \mdebye \sim g^3 \, T$
\cite{gava_jengo}.  The coefficient $e_2$ is positive,
so at asymptotically high temperatures,
the renormalized Polyakov loop
approaches unity from above.  
A related phenomenon is familiar in Coulombic plasmas.

Another case where a gauge field develops a mass is
when it is coupled to a scalar field, $\phi$, which then undergoes
spontaneous symmetry breaking.  
Then the gauge fields acquire a Higgs mass, $m_{\rm Higgs}$, and
there is a zero point energy $\sim g^2 m_{\rm Higgs}$.

The last example where a perturbative zero point energy arises is
for a gauge theory in three spacetime dimensions. Then the gauge coupling,
$g^2_{3d}$, has dimensions of mass, and there can be
a perturbative contribution to the
zero point energy $\sim g^2_{3 d}$.  This cannot be distinguished
from the nonperturbative contribution, since the square root of the
string tension is also $\sim g^2_{3 d}$.  In this sense, it is much
cleaner separating $\enon$ from $\epert$ in four, instead of three,
dimensions.

In fact, in three dimensions one cannot avoid a zero point
energy proportional to $g^2_{3 d}$ \cite{renloop_pert,ren_lat2}.
For a smooth Wilson loop in three dimensions, the only ultraviolet
divergence is at one loop order; instead of Eq.~(\ref{eq1}), it is
\begin{equation}
E_0^{{\rm pert} \, , \, 3d}
\sim -  \; g^2_{3d} \; \int^{1/a}_{1/L} \frac{d^2k}{(2\pi)^{2}} \;
\frac{1}{k^2} 
= - \; g^2_{3d} \;
c_2^{3 d} \; \log\left( \frac{L}{a} \right) \; .
\label{eq_3d}
\end{equation}
This then implies that bare loops vanish in the continuum limit as
\begin{equation}
{\cal Z}^{\rm div} _{{\cal R}}(g^2_{3d})
\sim \exp\left(-\; {\cal C}_{\cal R} c_2^{3 d} \; g^2_{3 d}  \; L
\; \log\left(L/a\right) \right) \sim
\left( \frac{a}{L}\right)^{{\cal C}_{\cal R} c_2^{3 d} \; g^2_{3 d}  \; L} \; .
\label{eq_3db}
\end{equation}
As in four dimensions, in three dimensions bare loops
vanish in the continuum limit, $a \rightarrow 0$.  The suppression
in three dimensions is only by a power of $a$, though, and is much
weaker than the exponential suppression seen in four dimensions.
This is seen in numerical simulations, where the bare Polyakov loop 
approaches unity at temperatures as low as several times 
the critical temperature \cite{petersen}.  

\section{String models}
\label{sec_string}

To epitomize how a nonperturbative zero point energy can arise,
in this section 
we consider a string model of the flux sheet.  Of course there are many other
models of confinement; presumably generic models
also generate $\enon \neq 0$.

We consider a string action
\begin{equation}
{\cal S} = \int d^2 z \; \sqrt{g} \;
g^{a b} \; \left( {\cal D}_a  \, x_\mu \right) \left(
\sigma + \frac{1}{\kappa} \; {\cal D}^2 
+ \frac{1}{\lambda} \; {\cal D}^4 \right) \left( {\cal D}_b \, x_\mu
\right) \; .
\label{string_action}
\end{equation}
Here ${\cal D}_a$ are covariant derivatives with respect to the 
induced metric $g_{a b} = \partial_a x^\mu \; \partial_b x^\mu$ on
the surface $x^{\mu}(z)$; 
${\cal D}^2 x^\mu = 1/\sqrt{g} \; \partial_a( \sqrt{g}  \; g^{a b} \;
\partial_b x^\mu)$.

In Eq.~(\ref{string_action}), $\sigma$ is the string tension, with
dimensions of mass squared.  The coupling for the extrinsic curvature
term, $\kappa$, is dimensionless.  Lastly, the coupling $\lambda$ also has
dimensions of mass squared.  This action can be considered as the
first terms in a power series of covariant derivatives on the world sheet.
It will be clear later that our qualitative conclusions will not be altered
by the presence of higher terms.

\subsection{Nambu model}

The simplest model is that for which $\kappa = \lambda = \infty$, so the
action only involves the area of the world sheet.  
In four dimensions, the exact solution for the potential is \cite{arvis}
\begin{equation}
V(R) = \sigma  \; \sqrt{ R^2 - \frac{\pi}{6\, \sigma } }
\; \approx \; \sigma R - \frac{\pi}{12 R} 
- \frac{\pi^2}{288 \, \sigma\, R^3} + \ldots \; .
\label{arvis}
\end{equation}
The term $\sim 1/R$ is universal \cite{lu_term,lu_wz,rigid2}.  
The potential is imaginary when  $R < \sqrt{\pi/(6 \sigma)}$, which reflects
the inconsistency of the pure Nambu model in four dimensions.

For our purposes, all we need to recognize is that at large
distances, the potential is $\sigma R$ times a power series in 
$1/(\sigma R^2)$.  Thus there is no zero point energy in the Nambu model,
$\enon = 0$.

We shall see in the next subsection that this arises because there are
only massless modes in the Nambu model.  Thus $\enon$ remains
zero for the models of Polchinski and Strassler \cite{pol_strom}
and of L\"uscher and Weisz \cite{lu_wz}.  While these models are
rather different, in both higher derivative terms,
like the 
extrinsic curvature term of Eq.~(\ref{string_action}), are treated as
perturbations to the Nambu model.  Thus there are only massless modes
in such models, and $\enon$ remains zero.

\subsection{Rigid strings}

We next consider the model with positive sign for the coupling to the
extrinsic curvature, $\kappa > 0$, and neglect the coupling for a 
higher derivative term, $\lambda = \infty$.  

We compute for small fluctuations about a flat sheet, and follow 
Alvarez \cite{alvarez}.  We let the number of spacetime dimensions, $d$,
be arbitrary.  At one loop order the $d-2$ transverse
fluctuations contribute to give an effective action,
\begin{equation}
S_\text{eff} =  \left( \frac{d-2}{2} \right) \; {\rm tr} \; \log
\left( -\partial^2 \right)
\left( -\partial^2 + m^2 \right) 
\;\; ; \;\;
- \partial^2 = -\partial_t^2 - \partial_r^2
\; \; , \; \; m^2 = \kappa \; \sigma \; .
\label{rig1}
\end{equation}
The time is continuous, so the corresponding momenta $\omega$ is
continuous.  In the spatial direction, the transverse directions vanish
at the end of the flux sheet, $x_{\rm tr}^\mu(0) = x_{\rm tr}^\mu(R) = 0$.
Thus the associated momenta are discrete, $p = n \pi/R$, for $n = 1, 2\ldots$.

The effective Lagrangian involves a product of a massless mode, from the
Nambu model, and a massive mode.  The integral over the massless mode
generates the term $\sim 1/R$.  To do the integral over the massive
mode, one can use analytic regularization \cite{alvarez}, or
consider the derivative with respect to $m^2$:
\begin{equation}
\frac{\partial S_\text{eff}}{\partial m^2}
= \left( \frac{d-2}{2} \right) 
\; {\rm tr} \; \frac{1}{- \partial^2 + m^2} 
= t_\text{tot} \; \left( \frac{d-2}{2} \right) 
\; \sum_{n=1}^\infty
\; \int \frac{d\omega}{2 \pi} \;
\frac{1}{\omega^2 + (n \pi/R)^2 + m^2} \; .
\label{rig2}
\end{equation}

The $\omega$ integral can be done either directly, or by contour integration:
\begin{equation}
\frac{\partial S_\text{eff}}{\partial m^2}
= t_\text{tot} \left( \frac{d-2}{4} \right) 
\; \sum_{n=1}^\infty
\; \frac{1}{\left(\omega^2 + (n \pi/R)^2 + m^2\right)^{1/2}} \; .
 \label{rig3}
\end{equation}
Integrating with respect to $m^2$, we obtain
\begin{equation}
S_\text{eff}(m^2)
= t_\text{tot} \left( \frac{d-2}{2} \right) 
\; \sum_{n=1}^\infty \; 
\left( \left( \frac{n \pi}{R}\right)^2 + m^2\right)^{1/2} \; .
\label{rig4}
\end{equation}
The sum over $n$ is highly divergent, but can be done by using
$\zeta$-function regularization.  

For $m=0$, there is a single term, $\sim 1/R$.  The sum over $n$ gives
$\zeta(-1)$, and generates the L\"uscher term.

When $m \neq 0$, we can expand the power series in powers of $n \pi/R$.
The first term is proportional to $m$ times $\zeta(0) = -1/2$, so that
the nonperturbative zero point energy is
\begin{equation}
\enon = - \left(\frac{d-2}{4}\right) \; \sqrt{\kappa \, \sigma} \; .
\label{rig5}
\end{equation}

This result was obtained previously \cite{rigid3,rigid4},
by computing in a large $d$ 
expansion in weak coupling, $\kappa \ll 1$.
Braaten, Pisarski, and Tse computed to leading order, and obtained
Eq.~(\ref{rig5}), Eq.~(16) of \cite{rigid3}.
Braaten and Tse then computed at next to leading order in $\kappa$,
and so determined the corrections $\sim \kappa \log(m)$ to Eq.~(\ref{rig5}),
Eq.~(4.24) of \cite{rigid4}.

Expanding $S_\text{eff}$ in powers of the spatial momenta, $(n \pi/R)^2$,
in principle we would expect corrections to the potential
proportional to $\sim 1/R^2$, 
$\sim 1/R^4$, and so on.  However, at one loop order a term
$\sim 1/(R^2)^\ell$ is proportional to $\zeta(-2\ell)$;
this vanishes when $\ell$ is an integer, as a 
``trivial'' zero of the $\zeta$-function.
Consequently, at one loop order there is {\it only} a contribution to
the zero point energy, as {\it all} other corrections to the potential
vanish.  It is amusing to contrast this to the massless case, where
simply on dimensional grounds the only
contribution to Eq.~(\ref{rig4}) is $\sim 1/R$.

That corrections $\sim 1/(R^2)^\ell$ vanish is 
special to one loop order.  This can be seen by the computations at
next to leading order in $\kappa$ at large $d$, where 
from Eq.~(4.23) of \cite{rigid4},
at large distances the potential is
\begin{equation}
V(R \rightarrow \infty) \; \approx \;
\sigma R - \frac{d}{4} \; m - \frac{\pi d}{24} \;\frac{1}{R}
- \frac{\pi^2  d^2}{1152} \; \frac{1}{\sigma R^3}
+ \frac{\pi^2 d^2}{384} \; \frac{1}{m \sigma R^4}
+ \ldots
\label{rig6}
\end{equation}
The result for the Nambu model is obtained by taking $d \rightarrow d-2$;
then for $d=4$,
the terms $\sim 1/R$ and $\sim 1/R^3$ agree with Eq.~(\ref{arvis}).
It is noteworthy that even at next to leading order in $\kappa$,
that there is no correction $\sim 1/R^2$.  We do not know if this
is peculiar to next to leading order, or persists at higher order.

There is a nonzero contribution $\sim 1/R^4$, which 
can be understood as follows.
At leading order, Eq.~(\ref{rig4}), there is 
a term $\sim 1/(m^3 R^4)$ times $\zeta(-4)$, which vanishes. 
At next to leading order there is a contribution
$\kappa$ times $\sim 1/(m^3 R^4)$, or
$\sim 1/(m \sigma R^4)$, which is nonzero.

\subsection{Confining strings}

We turn next to the case of confining strings, 
\cite{conf1,conf2,conf_temp}.  We take this to mean a
theory with negative coupling constant for the extrinsic curvature,
$\kappa < 0$, and $\lambda > 0$.  The effective Lagrangian is 
\begin{equation}
S_\text{eff} = \left( \frac{d-2}{2} \right) \; {\rm tr} \; \log
\left( -\partial^2 \right)
\left( (-\partial^2)^2 - 2 M_1^2 (-\partial^2) + M_2^4 \right)  \; ,
\label{conf1}
\end{equation}
where
\begin{equation}
2 M_1^2 = \frac{\lambda}{\kappa}
\;\;\; , \;\;\;
M_2^4 = \lambda \sigma \; .
\label{conf2}
\end{equation}
As usual, the massless mode generates the usual term $\sim 1/R$.  

The poles of the massive propagator are at
\begin{equation}
p^2 = M_1^2 \pm i \sqrt{M_2^4 - M_1^4} 
= M_2^2 \; {\rm e}^{\pm 2 i \theta} \;\; , \;\;
\tan(2 \theta) = \sqrt{\left( \frac{M_2}{M_1}\right)^4 - 1} \; .
\label{conf3}
\end{equation}
For the model to be physical, it is necessary that there are no poles on the
real axis \cite{conf2}.  This gives the constraint,
\begin{equation}
M_2^4 > M_1^4 \;\;\; , \;\;\;
\kappa^2 > \frac{\lambda}{4 \sigma} \; .
\label{conf4}
\end{equation}

As in the previous subsection we compute the derivative of the effective
Lagrangian with respect to $M_2^4$,
\begin{equation}
\frac{\partial S_\text{eff}}{\partial M_2^4}
= \left( \frac{d-2}{2} \right)\;
{\rm tr} \; \frac{1}{(-\partial^2)^2 - 2 M_1^2 (-\partial^2) + M_2^4}
\; .
\label{conf5}
\end{equation}
In taking this derivative we can assume that $M_1$ is independent of $M_2$.

We first compute the zero point energy, neglecting the dependence upon
the spatial momentum.  This implies that the sum over the discrete
spatial momenta generates $\zeta(0) = -1/2$.

It is then necessary to perform the integral over energies, $\omega$.
For rigid strings, the propagator has two poles, 
at $\pm i m$.  In the present case there are four poles.  Two are
in the upper half plane, $M_2 \, {\rm e}^{i \theta}$ and
$M_2 \, {\rm e}^{i(\pi - \theta)}$, and two in the lower half plane,
$M_2 \, {\rm e}^{- i \theta}$ and  $M_2 \, {\rm e}^{- i(\pi + \theta)}$.  
Including the contribution just of the two poles in the upper half plane,
the pole at $M_2 \, {\rm e}^{i \theta}$ gives a residue
$\sim {\rm e}^{- i \theta}/(M_2^3 \sin(2 \theta))$, while
that at $- M_2 \, {\rm e}^{-i \theta}$ gives a residue
$\sim {\rm e}^{+i \theta}/(M_2^3 \sin(2 \theta))$.  Hence,
\begin{equation}
\frac{\partial S_\text{eff}}{\partial M_2^4}
= -t_\text{tot} \left( \frac{d-2}{16} \right)\;
\frac{1}{\sin(\theta) \, M_2^3} \; .
\label{conf6}
\end{equation}
Using $\sin(\theta) = \sqrt{(1-M_1^2/M_2^2)/2}$, we find the
nonperturbative contribution to the zero point energy to be
\begin{equation}
\enon = - \left( \frac{d-2}{2 \sqrt{2}} \right)
\; \sqrt{M_2^2 - M_1^2} \; .
\label{conf7}
\end{equation}
This vanishes when $M_2 = M_1$, but at this point the propagator
has poles for real, 
Euclidean momenta, and the theory is not well defined \cite{conf2}.

Having obtained this result, one can also go through the algebra to
determine corrections to the potential involving higher powers of $1/R^2$.
When $p = n \pi/R$ is included, the poles in the propagator are shifted.
Even so, then only way that the spatial momentum $p$ enters is as $p^2$.
Thus if one expands the result in powers of $p$, one will find that
corrections are integral powers in $p^2$.  At one loop order, as for
the rigid string this only involves the trivial zeroes of the $\zeta$-function,
so that once again, the zero point energy is the {\it only} contribution
to one loop order.  As for the rigid string, we do not expect that this
persists to higher loop order.

Clearly one could consider higher derivative terms for an effective string
model.  By a similar analysis one expects that at one loop
order, the only correction to the Nambu potential is $\enon < 0$. 

\section{Conclusions}

We argued in Sec.~\ref{ren_wilson} that a renormalized quark antiquark
potential can be obtained from numerical simulations on the lattice,
and that perturbative contributions
to the associated zero point energy vanish.  
Nonperturbative contributions were computed in effective string
models in Sec.~\ref{sec_string}.  They vanish in the Nambu model, 
but arise for
either rigid or confining strings, with $\enon < 0$ at one loop order.

Numerical simulations appear to find that at large distances, corrections
to the Nambu model are small \cite{lattice_potential}.  This suggests
that on the world sheet, any massive modes are heavy on the scale of
$\sqrt{\sigma}$.  This is a regime of strong coupling in effective
string models, and so the one loop result for $\enon$ is not
definitive.  It will be interesting to see what numerical simulations find for 
both the sign and magnitude of $\enon$.

\section{Acknowledgements}
This research of R.D.P. was supported 
by the U.S. Department of Energy under
cooperative research agreement \#DE-AC02-98CH10886.
R.D.P. also thanks
the Alexander von Humboldt Foundation for their support.
This research of Y.H. was supported by the Grant-in-Aid for
the Global COE Program ``The Next Generation of Physics, 
Spun from Universality and Emergence'' from the Ministry of 
Education, Culture, Sports, Science and Technology (MEXT) of Japan.
We thank D. Antonov, M. Creutz, Z. Fodor, S. Gupta, K. H\"ubner, O. Kaczmarek,
F. Karsch, C. P. Korthals Altes, J. Kuti, 
M. Laine, P. Orland, P. Petreczky, C. Pica, E. Shuryak,
R. Venugopalan, and especially L. Yaffe for discussions and comments.


\begin{thebibliography}{999}
%
\bibitem{renloop_pert}
J.-L.~Gervais and A.~Neveu, \npb{163}{189}{1980};
A.~M.~Polyakov, \ibid{164}{171}{1980};
V.~S.~Dotsenko and S.~N.~Vergeles, \ibid{169}{527}{1980};
I.~Y.~Arefeva, \plb{93}{347}{1980};
R. A. Brandt, F. Neri, and M. Sato, \prd{24}{879}{1981};
R. A. Brandt, A. Gocksch, F. Neri, and M. Sato, \ibid{26}{3611}{1982}.
%
%
\bibitem{lu_term}
L.~Brink and H.~B.~Nielsen,
\plb{45}{332}{1973};
M.~Luscher, K.~Symanzik and P.~Weisz,
\npb{173}{365}{1980};
M.~L\"uscher,
\ibid{180}{317}{1981}.
%
\bibitem{lattice_potential}
G.~S.~Bali,
\phr{343}{1}{2001}
\href{http://arxiv.org/abs/hep-ph/0001312}{[arXiv:hep-ph/0001312]};
G.~S.~Bali {\it et al.}  [TXL Collaboration and T(X)L Collaboration],
\prd{62}{054503}{2000}
\href{http://arxiv.org/abs/hep-lat/0003012}{[arXiv:hep-lat/0003012]};
B.~Bolder {\it et al.},
\ibid{63}{074504}{2001}
\href{http://arxiv.org/abs/hep-lat/0005018}{[arXiv:hep-lat/0005018]};
S.~Necco and R.~Sommer,
\npb{622}{328}{2002}
\href{http://arxiv.org/abs/hep-lat/0108008}{[arXiv:hep-lat/0108008]};
M.~Luscher and P.~Weisz,
\jhep{0207}{049}{2002}
\href{http://arxiv.org/abs/hep-lat/0207003}{[arXiv:hep-lat/0207003]};
K.~J.~Juge, J.~Kuti and C.~Morningstar,
\prl{90}{161601}{2003}
\href{http://arxiv.org/abs/hep-lat/0207004}{[arXiv:hep-lat/0207004]};
P.~Majumdar,
\npb{664}{213}{2003}
\href{http://arxiv.org/abs/hep-lat/0211038}{[arXiv:hep-lat/0211038]};
M.~Caselle, M.~Hasenbusch and M.~Panero,
\jhep{0405}{032}{2004}
\href{http://arxiv.org/abs/hep-lat/0403004}{[arXiv:hep-lat/0403004]};
H.~Meyer and M.~Teper,
\ibid{0412}{031}{2004}
\href{http://arxiv.org/abs/hep-lat/0411039}{[arXiv:hep-lat/0411039]};
G.~S.~Bali,
Few Body Syst.\  {\bf 36}, 13 (2005)
\href{http://arxiv.org/abs/hep-ph/0411206}{[arXiv:hep-ph/0411206]};
M.~Caselle, M.~Hasenbusch and M.~Panero,
\jhep{0503}{026}{2005}
\href{http://arxiv.org/abs/hep-lat/0501027}{[arXiv:hep-lat/0501027]};
\ibid{0601}{076}{2006}
\href{http://arxiv.org/abs/hep-lat/0510107}{[arXiv:hep-lat/0510107]};
\ibid{0603}{084}{2006}
\href{http://arxiv.org/abs/hep-lat/0601023}{[arXiv:hep-lat/0601023]};
J.~Kuti,
PoS {\bf LAT2005}, 001 (2006)
\href{http://arxiv.org/abs/hep-lat/0511023}{[arXiv:hep-lat/0511023]};
N.~D.~Hari Dass and P.~Majumdar,
\jhep{0610}{020}{2006}
\href{http://arxiv.org/abs/hep-lat/0608024}{[arXiv:hep-lat/0608024]};
\plb{658}{273}{2008}
\href{http://arxiv.org/abs/hep-lat/0702019}{[arXiv:hep-lat/0702019]};
M.~Billo, M.~Caselle and L.~Ferro,
\jhep{0602}{070}{2006}
\href{http://arxiv.org/abs/hep-th/0601191}{[arXiv:hep-th/0601191]};
M.~Caselle, M.~Hasenbusch and M.~Panero,
\ibid{0709}{117}{2007}
\href{http://arxiv.org/abs/0707.0055}{[arXiv:0707.0055]};
P.~Giudice, F.~Gliozzi and S.~Lottini,
\ibid{0903}{104}{2009}
\href{http://arxiv.org/abs/0901.0748}{[arXiv:0901.0748]}.
%
\bibitem{lu_wz}
M.~L\"uscher and P.~Weisz,
\jhep{0407}{014}{2004}
\href{http://arxiv.org/abs/hep-th/0406205}{[arXiv:hep-th/0406205]}.
%
\bibitem{teper}
B.~Bringoltz and M.~Teper,
\plb{645}{383}{2007}
\href{http://arxiv.org/abs/hep-th/0611286}{[arXiv:hep-th/0611286]};
\ibid{663}{429}{2008}
\href{http://arxiv.org/abs/0802.1490}{[arXiv:0802.1490]};
A.~Athenodorou, B.~Bringoltz and M.~Teper,
\ibid{656}{132}{2007}
\href{http://arxiv.org/abs/0709.0693}{[arXiv:0709.0693]};
\href{http://arxiv.org/abs/0812.0334}{[arXiv:0812.0334]}.
%
\bibitem{ren_lat1}
O.~Kaczmarek, F.~Karsch, P.~Petreczky and F.~Zantow,
\plb{543}{41}{2002}
\href{http://arxiv.org/abs/hep-lat/0207002}{[arXiv:hep-lat/0207002]};
P.~Petreczky and K.~Petrov,
\prd{70}{054503}{2004}
\href{http://arxiv.org/abs/hep-lat/0405009}{[arXiv:hep-lat/0405009]};
O.~Kaczmarek, F.~Karsch, F.~Zantow and P.~Petreczky,
\ibid{70}{074505}{2004}
[Erratum-\ibid{72}{059903}{2005}]
\href{http://arxiv.org/abs/hep-lat/0406036}{[arXiv:hep-lat/0406036]};
M.~Doring, S.~Ejiri, O.~Kaczmarek, F.~Karsch and E.~Laermann,
\epjc{46}{179}{2006}
\href{http://arxiv.org/abs/hep-lat/0509001}{[arXiv:hep-lat/0509001]};
K.~Hubner, F.~Karsch, O.~Kaczmarek and O.~Vogt,
\prd{77}{074504}{2008}
\href{http://arxiv.org/abs/0710.5147}{[arXiv:0710.5147]}.
%
\bibitem{ren_lat2}
A.~Dumitru, Y.~Hatta, J.~Lenaghan, K.~Orginos and R.~D.~Pisarski,
\prd{70}{034511}{2004}
\href{http://arXiv.org/abs/hep-th/0311223}{[arXiv:hep-th/0311223]}.
%
\bibitem{ren_lat3}
Y.~Aoki, Z.~Fodor, S.~D.~Katz and K.~K.~Szabo,
\plb{643}{46}{2006}
\href{http://arxiv.org/abs/hep-lat/0609068}{[arXiv:hep-lat/0609068]}.
%
\bibitem{ren_lat4}
M.~Cheng {\it et al.},
\prd{77}{014511}{2008}
\href{http://arXiv.org/abs/0710.0354}{[arXiv:0710.0354]};
A.~Bazavov {\it et al.},
\href{http://arxiv.org/abs/0903.4379}{[arXiv:0903.4379]}.
%
\bibitem{ghk}
S.~Gupta, K.~H\"ubner and O.~Kaczmarek,
\prd{77}{034503}{2008}
\href{http://arXiv.org/abs/0711.2251}{[arXiv:0711.2251]}.
%
\bibitem{gava_jengo}
E.~Gava and R.~Jengo,
\plb{105}{285}{1981}.
%
\bibitem{heller}
U.~M.~Heller and F.~Karsch,
\npb{251}{254}{1985};
U. Heller, private communication.
%
\bibitem{lattice_effective}
L.~Dittmann, T.~Heinzl and A.~Wipf,
\jhep{0406}{005}{2004}
\href{http://arxiv.org/abs/hep-lat/0306032}{[arXiv:hep-lat/0306032]};
T.~Heinzl, T.~Kaestner and A.~Wipf,
\prd{72}{065005}{2005}
\href{http://arxiv.org/abs/hep-lat/0502013}{[arXiv:hep-lat/0502013]};
C.~Wozar, T.~Kaestner, A.~Wipf, T.~Heinzl and B.~Pozsgay,
\ibid{74}{114501}{2006}
\href{http://arxiv.org/abs/hep-lat/0605012}{[arXiv:hep-lat/0605012]};
C.~Wozar, T.~Kaestner, A.~Wipf and T.~Heinzl,
\ibid{76}{085004}{2007}
\href{http://arxiv.org/abs/0704.2570}{[arXiv:0704.2570]};
A.~Dumitru and D.~Smith,
\ibid{77}{094022}{2008}
\href{http://arxiv.org/abs/0711.0868}{[arXiv:0711.0868]};
A.~Velytsky,
\ibid{78}{034505}{2008}
\href{http://arxiv.org/abs/0805.4450}{[arXiv:0805.4450]};
C.~Wozar, T.~Kastner, B.~H.~Wellegehausen, A.~Wipf and T.~Heinzl
\href{http://arxiv.org/abs/0808.4046}{[arXiv:0808.4046]}.
%
\bibitem{loop_phen}
R.~D.~Pisarski,
\prd{62}{111501(R)}{2000}
\href{http://arxiv.org/abs/hep-ph/0006205}{[arXiv:hep-ph/0006205]};
A.~Dumitru and R.~D.~Pisarski,
\plb{504}{282}{2001}
\href{http://arxiv.org/abs/hep-ph/0010083}{[arXiv:hep-ph/0010083]};
\ibid{525}{95}{2002}
\href{http://arxiv.org/abs/hep-ph/0106176}{[arXiv:hep-ph/0106176]};
\prd{66}{096003}{2002}
\href{http://arxiv.org/abs/hep-ph/0204223}{[arXiv:hep-ph/0204223]};
R.~D.~Pisarski,
\ibid{74}{121703(R)}{2006}
\href{http://arXiv.org/abs/hep-ph/0608242}{[arXiv:hep-ph/0608242]};
Y.~Hidaka and R.~D.~Pisarski,
\prd{78}{071501(R)}{2008}
\href{http://arxiv.org/abs/0803.0453}{[arXiv:0803.0453]};
\prd{80}{036004}{2009}
\href{http://arxiv.org/abs/arXiv:0906.1751}{[arXiv:0906.1751]},
and manuscript in preparation.
%
%
\bibitem{alvarez}
O.~Alvarez,
\prd{24}{440}{1981}.
%
\bibitem{arvis}
J.~F.~Arvis,
\plb{127}{106}{1983}.
%
\bibitem{pol_strom}
J.~Polchinski and A.~Strominger,
\prl{67}{1681}{1991}.
%
\bibitem{drummond}
J.~M.~Drummond,
\href{http://arxiv.org/abs/hep-th/0411017}{[arXiv:hep-th/0411017]};
N.~D.~Hari Dass and P.~Matlock,
\href{http://arxiv.org/abs/hep-th/0606265}{[arXiv:hep-th/0606265]};
J.~M.~Drummond,
\href{http://arxiv.org/abs/hep-th/0608109}{[arXiv:hep-th/0608109]}.
%
xb%
\bibitem{rigid1}
H.~Kleinert,
\plb{174}{335}{1986};
A.~M.~Polyakov,
\npb{268}{406}{1986}.
%
\bibitem{rigid2}
E.~Braaten, S.~M.~Tse and R.~D.~Pisarski,
\prl{60}{2806}{1988}.
%
%
\bibitem{rigid3}
E.~Braaten, R.~D.~Pisarski and S.~M.~Tse,
\prl{58}{93}{1987} [Erratum-\ibid{59}{1870}{1987}].
%
\bibitem{rigid4}
E.~Braaten and S.~M.~Tse,
\prd{36}{3102}{1987}.
%
\bibitem{negative}
K.~I.~Maeda and N.~Turok,
\plb{202}{376}{1988};
R.~Gregory,
\ibid{206}{199}{1988};
P.~Orland,
\npb{428}{221}{1994}
\href{http://arxiv.org/abs/hep-th/9404140}{[arXiv:hep-th/9404140]};
M.~Sato and S.~Yahikozawa,
\npb{436}{100}{1995}
\href{http://arxiv.org/abs/hep-th/9406208}{[arXiv:hep-th/9406208]};
E.~T.~Akhmedov, M.~N.~Chernodub, M.~I.~Polikarpov and M.~A.~Zubkov,
\prd{53}{2087}{1996}
\href{http://arxiv.org/abs/hep-th/9505070}{[arXiv:hep-th/9505070]}.
%
\bibitem{conf1}
A.~M.~Polyakov,
\npb{486}{23}{1997}
\href{http://arxiv.org/abs/hep-th/9607049}{[arXiv:hep-th/9607049]}.
%
\bibitem{conf2}
M.~C.~Diamantini and C.~A.~Trugenberger,
\plb{421}{196}{1998}
\href{http://arxiv.org/abs/hep-th/9712008}{[arXiv:hep-th/9712008]};
M.~C.~Diamantini and C.~A.~Trugenberger,
\npb{531}{151}{1998}
\href{http://arxiv.org/abs/hep-th/9803046}{[arXiv:hep-th/9803046]};
D.~V.~Antonov,
\plb{427}{274}{1998}
\href{http://arxiv.org/abs/hep-th/9804016}{[arXiv:hep-th/9804016]};
M.~C.~Diamantini, H.~Kleinert and C.~A.~Trugenberger,
\prl{82}{267}{1999}
\href{http://arxiv.org/abs/hep-th/9810171}{[arXiv:hep-th/9810171]};
\plb{457}{87}{1999}
\href{http://arxiv.org/abs/hep-th/9903208}{[arXiv:hep-th/9903208]};
D.~Antonov,
\epl{52}{54}{2000}
\href{http://arxiv.org/abs/hep-th/0003043}{[arXiv:hep-th/0003043]};
\ibid{54}{715}{2001}
\href{http://arxiv.org/abs/hep-ph/0012009}{[arXiv:hep-ph/0012009]};
\plb{543}{53}{2002}
\href{http://arxiv.org/abs/hep-th/0207092}{[arXiv:hep-th/0207092]};
J.~Alfaro, L.~Balart, A.~A.~Andrianov and D.~Espriu,
\ijm{18}{2501}{2003}
\href{http://arxiv.org/abs/hep-th/0203215}{[arXiv:hep-th/0203215]};
D.~Antonov,
\jhep{0309}{005}{2003}
\href{http://arxiv.org/abs/hep-th/0304100}{[arXiv:hep-th/0304100]};
D.~Antonov and M.~C.~Diamantini,
\href{http://arxiv.org/abs/hep-th/0406272}{[arXiv:hep-th/0406272]};
O.~Aharony and E.~Karzbrun,
\href{http://arxiv.org/abs/0903.1927 }{[arXiv:0903.1927]}.
%
\bibitem{conf_temp}
J.~Polchinski and Z.~Yang,
\prd{46}{3667}{1992}
\href{http://arxiv.org/abs/hep-th/9205043}{[arXiv:hep-th/9205043]};
M.~C.~Diamantini and C.~A.~Trugenberger,
\prl{88}{251601}{2002}
\href{http://arxiv.org/abs/hep-th/0202178}{[arXiv:hep-th/0202178]};
M.~C.~Diamantini and C.~A.~Trugenberger,
\jhep{0204}{032}{2002}
\href{http://arxiv.org/abs/hep-th/0203053}{[arXiv:hep-th/0203053]}.
%
\bibitem{petersen}
B. Petersson, private communication.  See also
P.~Bialas, A.~Morel and B.~Petersson,
\npb{704}{208}{2005}
\href{http://arxiv.org/abs/hep-lat/0403027}{[arXiv:hep-lat/0403027]};
P.~Bialas, L.~Daniel, A.~Morel and B.~Petersson,
\ibid{807}{547}{2009}
\href{http://arxiv.org/abs/0807.0855}{[arXiv:0807.0855]}.
%
\end{thebibliography}
\end{document}